\documentclass{article}

\usepackage{amsmath, amssymb, a4wide}
\usepackage{txfonts}
\usepackage[UKenglish]{babel}
\usepackage[utf8]{inputenc}
\usepackage{color}
\usepackage{booktabs}
\usepackage{graphicx}

\newcommand{\VEC}[1]{\mathbf{#1}}

\newcommand{\elabel}[1]{\label{e:#1}}
\newcommand{\flabel}[1]{\label{f:#1}}

\newcommand{\eq}[1]{Eq.~\ref{e:#1}}

\newcommand{\fig}[1]{Supplementary Fig.~\ref{f:#1}}
\newcommand{\xvec}{\VEC{x}}

\usepackage[numbers]{natbib}
\begin{document}
\renewcommand{\labelenumi}{(\arabic{enumi})}
\title{Efimov-driven phase transitions of the unitary Bose gas\\
\emph{Supplementary information}}
\author{Swann Piatecki and Werner Krauth}
\date{\em Laboratoire de Physique Statistique, \'Ecole Normale Supérieure,
UPMC, Université Paris Diderot, CNRS, 24 rue Lhomond, 75005 Paris, France}

\maketitle

% \wk{nom, adresse, titre du papier, suivi par subtitle Supplementary information, etc}
% \sw{À voir les supplementary informations de quelques papiers, ils ne font pas ça.
% D'ailleurs, c'est eux qui demandent de ne pas mettre les numéros de page.}
% \pagestyle{empty}

% Redéfinition des indicateurs des figures demandée par Nature
\renewcommand{\figurename}{Supplementary Figure}
%\renewcommand\figurename{Supplementary Fig.}
% \renewcommand\thefigure{S\arabic{figure}}
% \renewcommand\thesection{S\arabic{section}}
%
% vector notations
%

\section{Model hamiltonian and unitary limit} 
\label{sec:hamiltonian-pi}

The pair interaction in the model hamiltonian may be viewed as the zero-range
limit of the square-well interaction potential shown in \fig{pair_potential}, 
whose range $r_0$ and depth $V_0$ are simultaneously taken to $0$ and $\infty$ 
so that the scattering length $a=r_0\left[1-\tan(r_0\sqrt{mV_0/\hbar^2})/
(r_0\sqrt{mV_0/\hbar^2})\right]$ remains constant. At zero range, the pair 
interaction only affects $s$-waves by the so-called Bethe-Peierls limit 
condition when the separation $r$ between two particles goes to zero. 

Unitarity corresponds to the pair interaction having a bound dimer of zero
energy and infinite extension, a limit in which the $k$-vector of the
particle is much smaller than $1/r_0$, and the scattering cross-section
$\sigma$ saturates the unitary limit $\sigma<8\pi/k^2$. This follows from the 
unitarity of the scattering operator.

\begin{figure}[htb]
  \centering
  \includegraphics{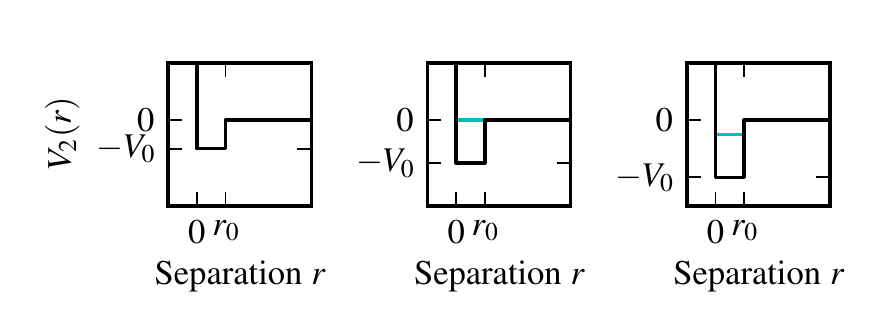}
  \caption{Illustration of square pair interaction potentials in the unbound region 
$R_0/a\sim -1$ \emph{(left)}, in the unitary region $R_0/a=0$ \emph{(centre)},
and in the bound pair + particle region $R_0/a\sim1$ \emph{(right)}. The zero-range
potential is fixed-$a$ limit of these potentials when $V_0\to\infty$ and $r_0\to0$.
The (possible) pair bound state correspond to cyan levels.}\flabel{potentials}
\flabel{pair_potential}
\end{figure}

While the two-particle properties are universal, the three-boson trimer ground
state  at unitarity generally depends on the details of the pair interaction. 
Excited trimers form a geometric sequence of asymptotically universal Efimov 
trimers with a fixed energetic ratio $E_{n}/E_{n+1}\underset{n\to\infty}
\rightarrow 515.03$, where $E_{n}$ is the energy of the $n$-th excited trimer. 
For the zero-range pair interaction and the three-body hard core,
the ground-state trimer is almost identical to universal Efimov trimers\cite{Braaten2006}.

\section{Dedicated Path-Integral Monte Carlo algorithm}

In our Path-Integral Monte Carlo algorithm, statistical weights due to 
the two-body and three-body interactions
in the model hamiltonian are computed as follows:
\begin{itemize}
\item the zero-range interaction is implemented  through
the \emph{pair-product approximation} for the density matrix, that estimates the
statistical weight of a configuration from the isolated two-body wavefunction of nearby
particles. For the zero-range interaction, only $s$-waves differ from the isolated system
of two non-interacting bosons. The pair-product approximation is valid when the 
imaginary time discretization step $\Delta\tau$ is small enough. 
\item The hyperradial cutoff $V_3$ is enforced by rejecting the
configurations where the hyperradius is smaller than a threshold $R_0$, an
approximation also valid if $\Delta\tau$ is small enough. In practice,
for the values of $\Delta\tau$ retained in our simulations, we need to
take in account a finite $\Delta\tau$ shift of the input value of $R_0$, which
we compute from a fit of the three-particle hyperradial probability distribution to
its analytically-known value\cite{Braaten2006},
\begin{equation*}
\elabel{p(R)}
  p(R)\propto R K^2_{is_0}\left(\sqrt{2}\kappa R\right), 
\end{equation*}
where $K_{is_0}$ a Bessel function of imaginary index $is_0\approx
1.00624i$, and $\kappa^2=mE_t/\hbar^2\propto R_0^{-2}$.
\end{itemize}

These weights set the probability of acceptance of the following
four types of Markov-chain moves, following a Metropolis-Hastings procedure.
\begin{enumerate}
\item Rebuild a single-particle path  on a fraction of the total imaginary
  time,
\item Move a whole permutation cycle,
\item Exchange two strands of paths  on a fraction of the
  total imaginary time,
\item Perform a compression-dilation move on one imaginary time slice 
(see \fig{comp_dil}).
\end{enumerate}

\begin{figure}[htb]
\centering
\includegraphics{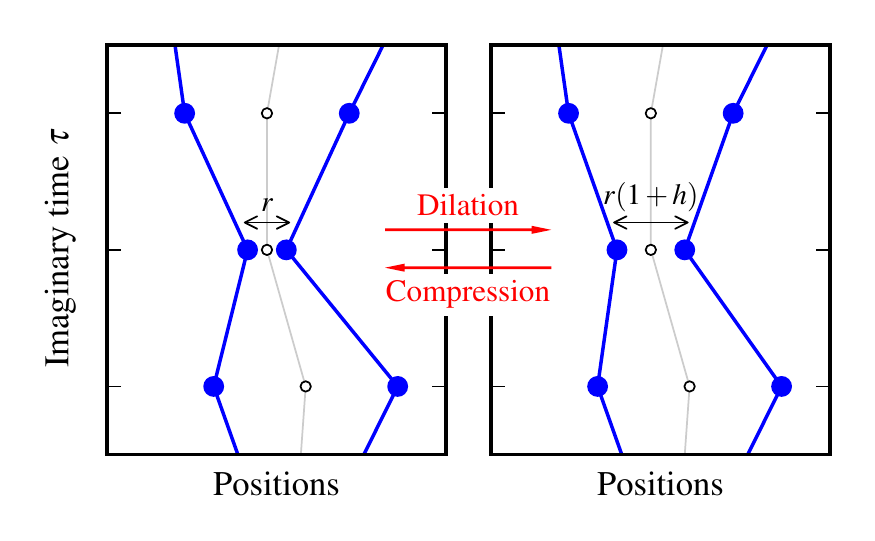}
\caption{Compression-dilation move (move (4)). At one imaginary time
  slice, the new proposed positions for two particles \emph{(blue
    paths)} are either more distant or closer by a factor $1+h$ while
  the centre of mass \emph{(white circles)} and positions at other time
  slices are kept in place. }
\flabel{comp_dil}
\end{figure}

Updates (1-3) are commonly used in Path-Integral Monte
Carlo simulations\cite{Pollock1984, Krauth1996, Krauth2006}. 
The compression-dilation move (4) specifically
addresses the divergence of the pair-distance
distribution function (see~Fig.1{\bf d}):
Even when it rebuilds a path on one single imaginary time slice, the
update (1) is linear: if two particles are separated by a distance
$r$, it will typically propose a new configuration with a separation
$r+\Delta$. The weight of the proposed configuration $w(r+\Delta)$ is
quite different from the weight of the original configuration $w(r)$
when $r\to0$, which results in a very low acceptance rate. The update
(4) proposes a configuration $r(1+h)$, whose weight $w(r) + hrw'(r)$
is close to $w(r)$: Because $w(r)$ diverges as $1/r^2$, $rw'(r)$ is of
the same order as $w(r)$, which generates much higher acceptance rates.

\section{Cocyclicity condition and graphic representation 
in Fig.1}

In our simulations of unitary bosons, the system is
contained in a harmonic trap, which regulates the available
configuration space. To reduce entropic effects for the three-body calculations
in Fig.1, we regulate the volume by imposing that they lie on
a single permutation cycle: In Fig.1{\bf a-c}, at imaginary time $\beta$, 
the blue, red and green bosons are respectively exchanged with the red, green and blue
bosons. This condition does not modify the properties of the fundamental
trimers as other permutations could be sampled at no cost at points of 
close encounters, some of which are highlighted in Fig.1{\bf b}.

In Fig.1{\bf a-c}, four-dimensional co-cyclic path-integral configurations
are represented in a three-dimensional plot. As the motion of the
centre of mass is decoupled from the effects of the interactions, its
position is set to zero at all $\tau$. The three
spatial dimensions are then reduced to two dimensions by rotating the
triangle formed by the three particles at each imaginary time to the
same plane in a way that does not favour any of the three spatial
dimensions, but that conserves the permutation cycle structure. This
transformation conserves the pair distances.

\section{High-temperature equation of state}

The local density approximation consists in assuming that, at each
point $\xvec= (x,y,z)$ of the trap, with $|\xvec| = r$, the Bose gas is in
equilibrium at a chemical potential $\mu=\mu_0-\omega^2 r^2/2$, where $\mu_0$
is the chemical potential in the trap centre. Within this
approximation, thermodynamical identities allow one to 
relate the doubly-integrated density profile $\bar n(x)$,
and the grand-canonical
pressure\cite{Ho2010}:
\begin{equation}
  P(\mu(x)) = \frac{\omega^2}{2\pi}\bar{n}(x),
\end{equation}
where $\mu(x)=\mu_0-\omega^2x^2/2$. The 
chemical potential $\mu_0$  at the centre of the trap can be obtained 
from a fit in the outer region of the trap where $P(\mu)$ is the pressure
of a classical ideal gas.

The doubly-integrated density profile is obtained by ensemble averaging, 
and the numerical equation of state $P(\mu)$ is compared to the
expansion of the pressure in powers of the fugacity $e^{\beta\mu}$ (see Fig.3),
\begin{equation}
  P=\frac{k_BT}{\lambda^3}\sum_{l\geq 1}b_le^{l\beta\mu}.
\end{equation}

The $l$-th \emph{cluster integral} $b_l$
corresponds to the $l$-body effects that cannot be reduced to smaller
non-interacting groups of interacting particles\cite{Huang1987}. The
classical ideal gas corresponds to $b_1=1$, and higher coefficients
both describe both energetic and quantum-statistical correlations. The
$l$-th cluster integrals are related to the virial coefficients
$\{a_{l'}\}_{l'\leq l}$\cite{Huang1987}
as 
\begin{equation}
  \begin{cases}
    b_1=a_1=1,\\
    b_2=-a_2,\\
    b_3=-\frac{1}{2}a_3 + 2a_2^2.
  \end{cases}
\end{equation}

At unitarity, an analytical expression of the coefficients $b_2$ and
$b_3$ was obtained recently\cite{Castin2013}, $b_2=9/(4\sqrt{2})\approx1.59$,
 and
\begin{equation}
  \frac{b_3}{3\sqrt{3}} = C + \sum_{q\geq 0}\left(e^{-\beta \epsilon_q} - 1\right)
  + \frac{|s_0|}{\pi}\left\{\frac{1}{2}\ln\left(e^\gamma \beta E_t\right) - 
    \sum_{p\geq 1}e^{-p\pi |s_0|}\Re\left[\Gamma(-ip|s_0|)(\beta E_t)^{ip|s_0|}
\right]\right\},
\elabel{b3}
\end{equation}
where $C=0.648$ and $|s_0|=1.00624$ are two constants, $\gamma=0.577$
is the Euler constant, $\epsilon_q=-E_te^{-2\pi q/|s_0|}$ represents the
energy levels of the three-body bound states, and $-E_t$ is the energy
of the fundamental Efimov trimer, which is related to the 
average squared hyperradius in the fundamental state\cite{Braaten2006}:
\begin{equation}
  E_t = \frac{2(1+s_0^2)}{3\left\langle R^2\right\rangle},
\elabel{E_t}
\end{equation}
an expression that may be used to obtain 
$E_t\approx 4.27e^{-3}\hbar^2 /(m R_0^{2})$ by numerically integrating 
\eq{p(R)}.

In the temperature range of our simulations, it is sufficient to perform the 
sums in \eq{b3} up to $q=1$. 

\section{Monitoring the phase transition}

When the densities of the gas and the superfluid Efimov liquid approach
each other, observing directly the two-dimensional
histogram of pair distances and centre-of-mass positions does not allow to
distinguish between a
weakly first-order phase transition and a cross-over (see
\fig{joint_distdistrib}). 

\begin{figure}[htb]
\centering
\includegraphics{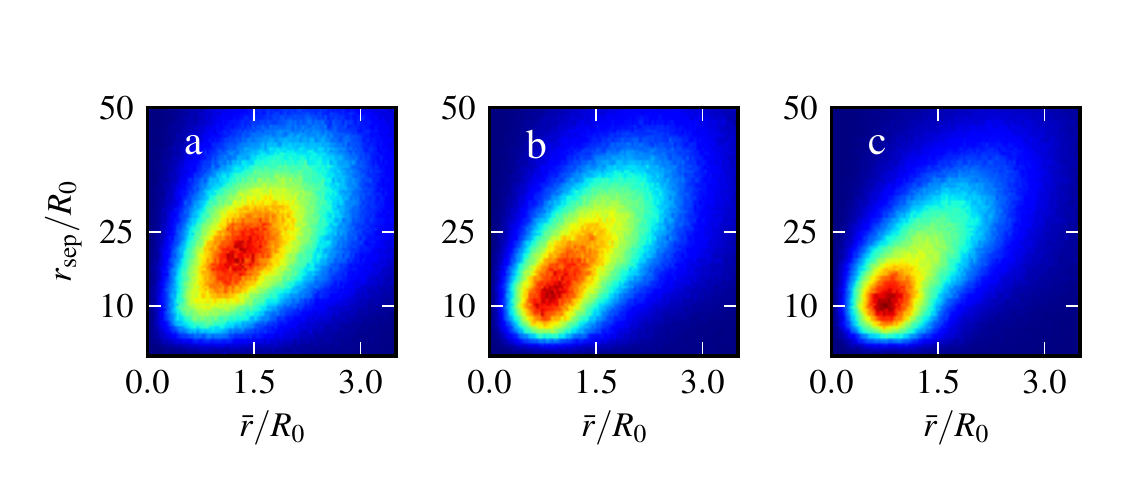}
\caption{Two-dimensional histogram of pair distances $r_\text{sep}$ 
  and centre-of-mass positions $\bar r$ for $R_0=1.3$,
  at $k_BT/\hbar\omega=4.3.$ (a), $k_BT/\hbar\omega=4.2$ (b), and
  $k_BT/\hbar\omega=4.0$. The sole observation of this histogram 
does not allow here to state that
  the system is undergoing a phase transition as the densities of
  both phases are too close.}
\flabel{joint_distdistrib}
\end{figure}
In this regime, we monitor the normal-gas-to-superfluid-liquid phase
transition more accurately by following the evolution of the first
peak of the pair correlation function (obtained by ensemble
averaging) with temperature (see \fig{knee}).

\begin{figure}[htb]
\centering
\includegraphics{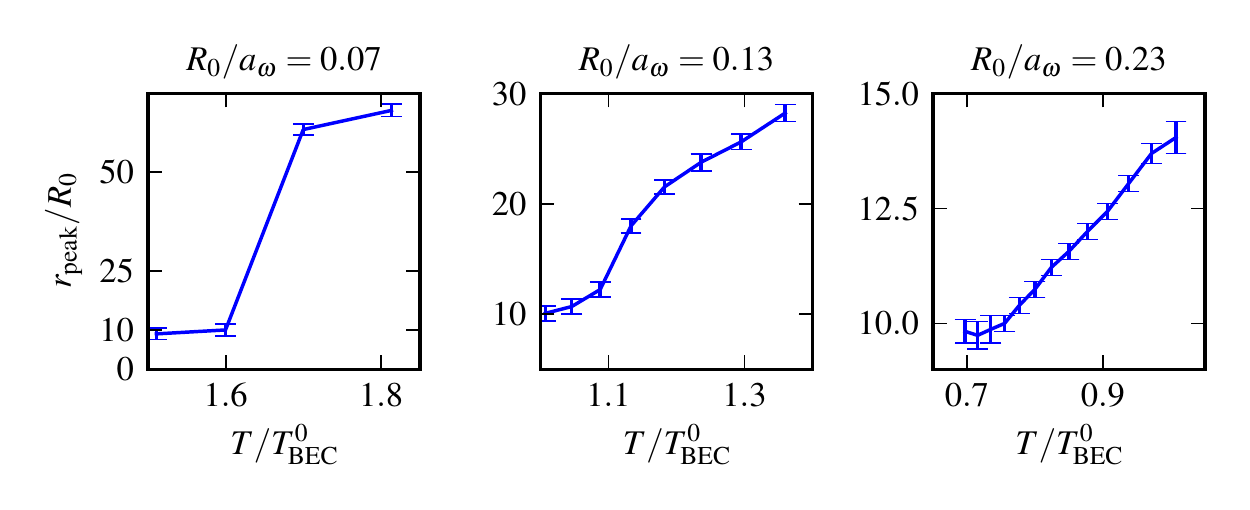}
\caption{Evolution of the position of the first peak of the pair
  correlation function for several values of $R_0$. For
  $R_0/a_\omega=0.07$, the large step indicates the first
  order phase transition (this corresponds to Fig.3{\bf a-c}); at
  $R_0/a_\omega=0.13$, the step persists but is considerably smaller,
  a sign that the system also undergoes a first order phase
  transition. At $R_0/a_\omega=0.23$, there is no sign of a first
  order phase transition.}\flabel{knee}
\end{figure}

For each value of $R_0$, simulations in the trap are run at a 
discrete set of inverse temperature $\beta$. In Fig.4{\bf a}, the error bars
in the normal-gas-to-superfluid-liquid transition line show the 
interval between the last temperature at
which there is no liquid droplet, and the first temperature at which 
there is one. The errors for the conventional superfluid transition are computed
in the same way, with the criterion that the gas is superfluid if
particles lie in a permutation cycle of length at least ten
with a probability higher than~$0.05$.

\section{Approximate semi-analytical phase diagram}

When the free energy $F$ of a homogeneous physical system is not a
convex function of its volume $V$, it becomes more favourable to split
the system into two phases than to keep the system homogeneous, a
situation from which first order phase transitions originate (see
\fig{free_energy}), and that yields the equality of the pressures and of the
chemical potentials of both phases at coexistence in absence of
interface energy\cite{Landau1980}.

\begin{figure}[htb]
  \centering
  \includegraphics{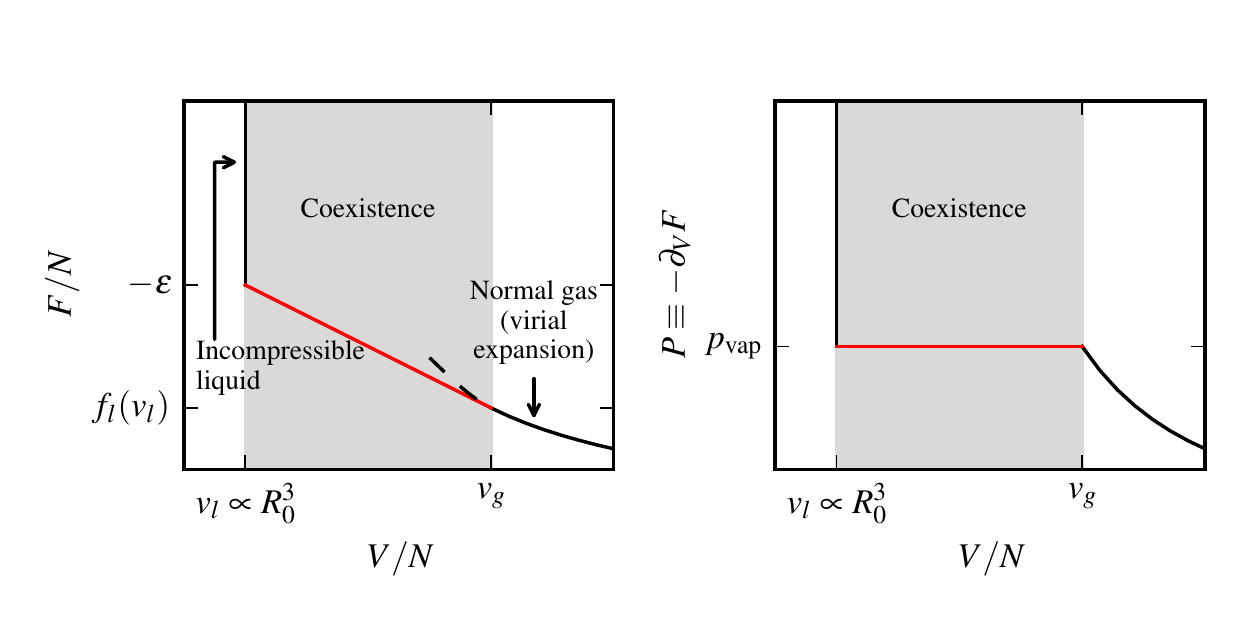}
  \caption{Sketch of the behaviour of the free energy and the pressure
    of unitary bosons through the transition in our incompressible
    fluid model.}
    \flabel{free_energy}
\end{figure}

In the system of unitary bosons, the virial expansion is an excellent
approximation for the normal gas phase far from the superfluid
transition. Although it becomes irrelevant in the superfluid gas, its
analytic continuation conveys important qualitative features, because
of the continuous nature of the normal-gas-to-superfluid-gas phase
transition, and is therefore a suitable poor man's approximation to the
behaviour of the superfluid gas.

The simplest theoretical model for the superfluid liquid is that of an
incompressible liquid of specific volume $v_l$ and negligible
entropic contribution to the free energy. The incompressibility
approximation is acceptable for unitary bosons as the first peak of
the pair correlation function seems to scale only with $R_0$ (see
Fig.2); simulations at high $R_0$ yield $v_l\approx(6R_0)^3$. The
negligibility of the entropic contribution to the free energy is
ensured for non-pathological systems at low temperature. As results of
Ref.\cite{Stecher2010} may be extrapolated to obtain an energy per
particle $-\epsilon=-10.1E_t$ in the liquid phase (see
\fig{vonStecher}), in this approximation, the free energy is
$F_l=-N\epsilon$.

\begin{figure}[htb]
  \centering
  \includegraphics{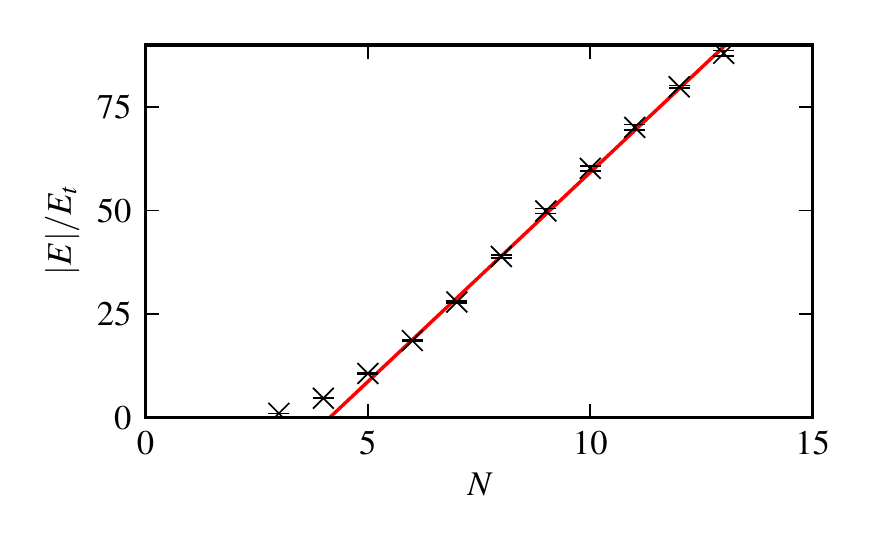}
  \caption{Energy of clusters of unitary bosons at zero temperature 
    \emph{(black crosses)} as a function of the number $N$ of
    bosons in the cluster, computed by J. von Stecher in
    Ref.\cite{Stecher2010}, in units of the fundamental Efimov
    trimer energy $E_t$. A linear regression \emph{(solid red line)}
    gives $|E|/N=10.1E_t$ for large $N$ values. For the phase diagram of $100$ bosons in a
    harmonic trap (Fig.4{\bf a}), the value of $|E|/N$ is empirically adjusted to
    $8E_t$ to account for the small number of bosons in the liquid droplet.}
    \flabel{vonStecher}
\end{figure}

In practice, we draw the transition line into the superfluid liquid by
finding the smallest chemical potential at which the pressures of the
incompressible liquid and of the gas phase (approximated by the
virial expansion) coincide:
\begin{equation}
  \frac{\mu-\epsilon}{v_l}=\frac{k_BT}{\lambda^3}\left(e^{\beta\mu}+b_2e^{2\beta\mu}+
b_3e^{3\beta\mu}\right).
\end{equation}

As $1/v=\partial_\mu P$, the crossing to the regime where this
equation has no solution corresponds to the critical point, where both
densities are equal.

To draw the coexistence line for the trap  centre with
$N=100$ particles, the chemical potential $\mu_0$ at the centre of the
trap is found from integrating the density throughout the trap:
\begin{equation}
  \frac{4\pi}{\lambda^3}\int_0^\infty r^2dr\left(e^{\beta\mu(r)}+2b_2e^{2\beta\mu(r)}+
3b_3e^{3\beta\mu(r)}\right) = N,
\end{equation}
where $\mu(r)=\mu_0-\omega^2r^2/2$.

\bibliography{../efimov_nature.bib}

\begin{thebibliography}{1}
\expandafter\ifx\csname url\endcsname\relax
  \def\url#1{\texttt{#1}}\fi
\expandafter\ifx\csname urlprefix\endcsname\relax\def\urlprefix{URL }\fi
\providecommand{\bibinfo}[2]{#2}
\providecommand{\eprint}[2][]{\url{#2}}

\bibitem{Braaten2006}
\bibinfo{author}{Braaten, E.} \& \bibinfo{author}{Hammer, H.-W.}
\newblock \bibinfo{title}{Universality in few-body systems with large
  scattering length}.
\newblock \emph{\bibinfo{journal}{Phys. Rep.}} \textbf{\bibinfo{volume}{428}},
  \bibinfo{pages}{259--390} (\bibinfo{year}{2006}).

\bibitem{Pollock1984}
\bibinfo{author}{Pollock, E.~L.} \& \bibinfo{author}{Ceperley, D.~M.}
\newblock \bibinfo{title}{Simulation of quantum many-body systems by
  path-integral methods}.
\newblock \emph{\bibinfo{journal}{Phys. Rev. B}} \textbf{\bibinfo{volume}{30}},
  \bibinfo{pages}{2555--2568} (\bibinfo{year}{1984}).

\bibitem{Krauth1996}
\bibinfo{author}{Krauth, W.}
\newblock \bibinfo{title}{Quantum {Monte Carlo} calculations for a large number
  of bosons in a harmonic trap}.
\newblock \emph{\bibinfo{journal}{Phys. Rev. Lett.}}
  \textbf{\bibinfo{volume}{77}}, \bibinfo{pages}{3695--3699}
  (\bibinfo{year}{1996}).

\bibitem{Krauth2006}
\bibinfo{author}{Krauth, W.}
\newblock \emph{\bibinfo{title}{Statistical Mechanics: Algorithms and
  Computations}} (\bibinfo{publisher}{Oxford University Press, {Oxford, Great
  Britain}}, \bibinfo{year}{2006}).

\bibitem{Ho2010}
\bibinfo{author}{Ho, T.-L.} \& \bibinfo{author}{Zhou, Q.}
\newblock \bibinfo{title}{Obtaining the phase diagram and thermodynamic
  quantities of bulk systems from the densities of trapped gases}.
\newblock \emph{\bibinfo{journal}{Nature Phys.}} \textbf{\bibinfo{volume}{6}},
  \bibinfo{pages}{131--134} (\bibinfo{year}{2010}).

\bibitem{Huang1987}
\bibinfo{author}{Huang, K.}
\newblock \emph{\bibinfo{title}{Statistical mechanics}}
  (\bibinfo{publisher}{Wiley}, \bibinfo{address}{New York},
  \bibinfo{year}{1987}), \bibinfo{edition}{2nd} edn.

\bibitem{Castin2013}
\bibinfo{author}{Castin, Y.} \& \bibinfo{author}{Werner, F.}
\newblock \bibinfo{title}{Le troisième coefficient du viriel du gaz de bose
  unitaire}.
\newblock \emph{\bibinfo{journal}{Canad. J. Phys.}}
  \textbf{\bibinfo{volume}{91}}, \bibinfo{pages}{382--389}
  (\bibinfo{year}{2013}).
\newblock \eprint{arxiv:1212/5512 (English version)}.

\bibitem{Landau1980}
\bibinfo{author}{Landau, L.~D.} \& \bibinfo{author}{Lifshitz, L.~M.}
\newblock \emph{\bibinfo{title}{Statistical physics}}.
\newblock No.~\bibinfo{number}{5} in \bibinfo{series}{Course of theoretical
  physics} (\bibinfo{publisher}{Butterworth-Heinemann}, \bibinfo{year}{1980}),
  \bibinfo{edition}{3rd} edn.

\bibitem{Stecher2010}
\bibinfo{author}{von Stecher, J.}
\newblock \bibinfo{title}{Weakly bound cluster states of {Efimov} character}.
\newblock \emph{\bibinfo{journal}{J. Phys. B}} \textbf{\bibinfo{volume}{43}},
  \bibinfo{pages}{101002} (\bibinfo{year}{2010}).

\end{thebibliography}

\end{document}